\documentclass[useAMS,usenatbib]{mn2e}
\usepackage{url}
\usepackage{graphicx}
\usepackage{verbatim}
\usepackage[usenames]{color}
\usepackage[flushleft]{threeparttable}
\DeclareGraphicsExtensions{.pdf,.png,.jpg,.mps,.eps,.ps}
\usepackage{amsmath}
\usepackage{natbib}
\usepackage{color}
\usepackage{graphicx}
\usepackage{lscape}
\usepackage{gensymb}

\newcommand\nustar{{\it NuSTAR}}

\newcommand\asca{{\it ASCA}}

\newcommand\sax{{\it BeppoSAX}}
\newcommand\chandra{{\it Chandra}}

\newcommand\suzaku{{\it Suzaku}}
\newcommand\rxte{{\it RXTE}}
\newcommand\xmm{{\it XMM-Newton}}

\newcommand\kev{{\rm~keV}}

\newcommand\kms{\ifmmode {\rm~km\ s}^{-1} \else ~km s$^{-1}$\fi}
\newcommand\Hunit{\ifmmode {\rm~km\ s}^{-1}\ {\rm Mpc}^{-1}
        \else ~km s$^{-1}$ Mpc$^{-1}$\fi}
\newcommand\ctssec{\ifmmode {\rm~count\ s}^{-1} \else ~count s$^{-1}$\fi}
\newcommand\ergsec{\ifmmode {\rm~erg\ s}^{-1} \else
        ~erg s$^{-1}$\fi}
\newcommand\funit{\ifmmode {\rm~erg\ s}^{-1}\;{\rm cm}^{-2} \else
        ~ergs s$^{-1}$ cm$^{-2}$\fi}
\newcommand\phflux{\ifmmode {\rm~photon\ s}^{-1}\;{\rm cm}^{-2}
        \else   ~photon s$^{-1}$ cm$^{-2}$\fi}
\newcommand\efluxA{\ifmmode {\rm~erg\ s}^{-1}\;{\rm cm}^{-2}\;{\rm
        \AA}^{-1} \else ~erg s$^{-1}$ cm$^{-2}$ \AA$^{-1}$\fi}
\newcommand\efluxHz{\ifmmode {\rm~erg\ s}^{-1}\;{\rm cm}^{-2}\;{\rm
        Hz}^{-1} \else ~erg s$^{-1}$ cm$^{-2}$ Hz$^{-1}$\fi}
\newcommand\cc{\ifmmode {\rm~cm}^{-3} \else cm$^{-3}$\fi}
\newcommand\FWHM{\ifmmode {\rm~FWHM} \else ${\rm~FWHM}$\fi}
\newcommand\Msun{\ifmmode M_{\odot} \else $M_{\odot}$\fi}
\newcommand\Lsun{\ifmmode L_{\odot} \else $L_{\odot}$\fi}

\newcommand\hbeta{\ifmmode {\rm H}\beta \else H$\beta$\fi}
\newcommand\Kalpha{\ifmmode {\rm K}\alpha \else K$\alpha$\fi}
\newcommand\nh{\ifmmode N_{\rm H} \else N$_{\rm H}$\fi}

\usepackage{graphicx}
\usepackage{url}
\usepackage{verbatim}
\usepackage{color}

\title[\nustar{} view of Ser~X-1]{On the disc reflection spectroscopy of NS LMXB Serpens~X-1: analysis of a recent \nustar{} observation}
\author[Mondal et al.]{\parbox[]{6.5in}{{Aditya S. Mondal$^{1}\thanks{E-mail: adityas.mondal@visva-bharati.ac.in}$, G. C. Dewangan$^{2}$
, B. Raychaudhuri$^{1}$ }  \\
\small
$^{1}$Department of physics, Visva-Bharati, Santiniketan, West Bengal-731235, India \\
$^{2}$Inter-University Centre for  Astronomy \& Astrophysics (IUCAA), Pune, 411007 India \\
}}

\date{\today}
\begin{document}
\maketitle
\begin{abstract}
We present \nustar{} observation of the atoll type neutron star (NS) low-mass X-ray binary (LMXB) Serpens~X-1 (Ser~X-1) performed on 17 February 2018. We observed Ser~X-1 in a soft X-ray spectral state with $3-79\kev{}$ luminosity of $L_\text{X}\sim0.4\times 10^{38}$ erg~s$^{-1}$ ($\sim 23\%$ of the Eddington luminosity), assuming a distance of 7.7 kpc. A positive correlation between intensity and hardness ratio suggests that the source was in the banana branch during this observation. The broadband $3-30\kev$ \nustar{} energy spectrum can be well described either by a three-component continuum model consisting of a disk blackbody, a single temperature blackbody and a power-law or by a two-component continuum model consisting of a disk blackbody and a Comptonization component. A broad iron line $\sim 5-8$ keV and the Compton back-scattering hump peaking at $\sim10-20 \kev{}$ band are clearly detected in the X-ray spectrum. These features are best interpreted by a self-consistent relativistic reflection model. Fits with relativistically blurred disc reflection model suggests that the inner disc radius $R_{in}$ is truncated prior to the ISCO at $(1.9-2.5)\;R_{ISCO}$ ($\simeq11.4-15\,R_{g}\: \text{or}\: 26-34$ km) and the accretion disc is viewed at an low inclination of $i\simeq16\degr-20\degr$. The disc is likely to be truncated either by a boundary layer or by the magnetosphere. Based on the measured flux and the mass accretion rate, the maximum radial extension for the boundary layer is estimated to be $\sim6.4\:R_{g}$ from the NS surface. The truncated inner disc in association with pressure from a magnetic field sets an upper limit of $B\leq1.9\times10^{9}$ G.
\end{abstract}
\section{introduction}
A Neutron Star Low Mass X-ray Binary (NS LMXB) is a compact system composed of an NS and a low-mass ($\leq1\:M_{\odot}$) companion star. NS LMXBs are classified into two main groups based on their X-ray luminosity along with the spectral and the timing properties in X-rays \citep{1989A&A...225...79H}. Those are the so-called ``Z'' sources, with luminosities close to or above the Eddington luminosity ($L_\text{Edd}$) and the ``atoll'' sources, with luminosities up to $\sim0.5\:L_\text{Edd}$ \citep{2010ApJ...719..201H}. The names of the Z and the atoll sources are related to the shape traced in the color-color diagram (CD). The Z sources show three-branches (the horizontal, the normal and the flaring branches) whereas the atoll sources show two main regions in the CD, the island state and isolated from it, the so-called banana branch. The X-ray spectra of the Z sources are soft in all branches and those of the atoll sources are soft at high luminosities and hard at low luminosities. The harder one is related to the island state and the softer one is related to the banana state which can be further divided into lower banana and upper banana states. However, the relation between the atoll and the Z-track sources is not well understood. Our understanding has improved with the discovery of the transient source XTE J1701-462 \citep{2006ATel..696....1R} which shows all the characteristics of a Z source as well as an atoll source during the decaying phase of the outburst \citep{2010ApJ...719..201H}. This implies that whether an NS is an atoll or a Z type is mainly determined by the mass accretion rate \citep{2007ApJ...667.1073L}. \\

In NS LMXBs, a geometrically thin, optically thick accretion disc is formed around the NS when it accretes matter from the companion star \citep{1973A&A....24..337S}. The radiation from an accretion disc generates a quasi-thermal spectrum. The accretion discs are usually accompanied by a hot corona \citep{1973A&A....24..337S} and the coronal emission of such source generates a (cutoff) power-law spectrum by inverse Compton scattering of the thermal disc photon. It is well known that the spectrum from an accretion disc is a multicolor blackbody. At the same time, another hot single-temperature blackbody may potentially arise due to the emission from the boundary layer between the inner accretion disc and the NS surface. This hard X-ray emission (either a cutoff power-law in the hard state or a blackbody component in the soft state) can irradiate the accretion disc to produce a reflection spectrum. High energy photons tend to Compton scatter back out of the disc, resulting in a broad hump-like shape in the reflection spectrum \citep{2001MNRAS.327...10B, 2007MNRAS.381.1697R}. In addition, several narrow emission lines are produced among which Fe K$\alpha$ fluorescent line is the most prominent one because of its high fluorescent yield and large cosmic abundance \citep{2007ApJ...664L.103B, 2008ApJ...674..415C, 2008ApJ...688.1288P, 2009MNRAS.399L...1R, 2015MNRAS.451L..85D}. The intrinsically narrow Fe K$\alpha$ lines when appear in the X-ray spectra of LMXBs show a broad, asymmetric profile due to Doppler and gravitational shift \citep{2000PASP..112.1145F}. Studies of Fe K$\alpha$ line profile provide an independent view of the inner accretion flow in NS LMXBs which led to constraints to the inner disc structure and inclination. The accretion disc in NS systems could be truncated by the boundary layer between the disc and the NS surface or by a strong stellar magnetic field. Thus the inner disc radius may give an upper limit to the radius of the NS and hence can constrain the NS equation of state \citep{2000A&A...360L..35P, 2008ApJ...674..415C, 2011MNRAS.415.3247B}. Alternatively, Fe K$\alpha$ line profiles can also be used to obtain an upper limit on the strength of the magnetic field associated with the NS \citep{2019ApJ...873...99L, 2016MNRAS.461.4049D, 2016ApJ...819L..29K}. \\

The bright persistent atoll type NS LMXB Ser~X-1 was discovered in 1965 \citep{1965Sci...147..394B}. Type-1 thermonuclear X-ray bursts have been detected from the source in 1976 \citep{1976IAUC.2963....1S, 1977MNRAS.179P..21L} and it confirms that the compact object in this source is an NS. A super-burst with a duration of approximately 4-hours has also been reported \citep{2002A&A...382..174C}. The source is located at a distance of $7.7\pm0.9$ kpc \citep{2008ApJS..179..360G}. The counterpart of Ser~X-1 was identified with a main-sequence K-dwarf star \citep{2013MNRAS.432.1361C}. After its discovery, it has been observed with all major X-ray missions like \asca{} \citep{2001A&A...369..915C}, \sax{} \citep{2001A&A...366..138O}, \rxte{} \citep{2001A&A...366..138O}, \xmm{} \citep{2007ApJ...664L.103B}, \suzaku{} \citep{2010ApJ...720..205C, 2016ApJ...831...45C}, \chandra{} \citep{2016ApJ...821..105C} and \nustar{} \citep{2013ApJ...779L...2M, 2017A&A...600A..24M}. Several continuum models were used in the previous works, using different combinations of disc blackbody, single temperature blackbody and power-law. Broadband observations preferred a continuum model consisting of all three components \citep{2008ApJ...674..415C, 2010ApJ...720..205C, 2013ApJ...779L...2M, 2016ApJ...831...45C}. On the contrary, observations with limited energy ranges preferred two component continuum models, either a disc blackbody and a single temperature blackbody or a single temperature blackbody and a power-law \citep{2007ApJ...664L.103B, 2016ApJ...821..105C}. There are little spectral changes in this source as it is usually found in the soft state. Relativistic broad iron lines have been reported from almost all the previous observations \citep{2007ApJ...664L.103B, 2010ApJ...720..205C, 2013ApJ...779L...2M, 2016ApJ...831...45C}. Different self-consistent reflection models have been used to fit the reflection component. \citet{2017A&A...600A..24M} stated that different results in different observations are probably due to different modelling of the continuum and/or the reflection component. Optical spectroscopy and some other X-ray reflection studies  point towards a low binary inclination $\sim10\degr$ \citep{2013ApJ...779L...2M, 2013MNRAS.432.1361C}, although some studies reported the inclination angle in between $25\degr-50\degr$ \citep{2016ApJ...821..105C, 2017A&A...600A..24M, 2010ApJ...720..205C}. Ser~X-1 gives us an oppurtunity to detect multiple reflection features because of the low-amount of absorping material in the line of sight of Ser~X-1 (corresponding to low neutral hydrogen column density $N_{H}=4.0\times10^{21}$ cm$^{-2}$ \citealt{1990ARA&A..28..215D}).  \\

In the present work, we analyze the latest $\sim31$ ks pile-up free \nustar{} observation of Ser~X-1 with the main aim to study the reflection component and put constraints on the inner disk parameters. This observation also allows us to study the source broadband spectrum which is also important to constrain the reflection component such as the broad Fe emission line along with the Compton hump. In this \nustar{} observation, the source was captured with a lower luminosity compared to the previous \nustar{} observation performed in 2013 ($\sim2$ times more luminous than the present observation). So, this observation can be useful to test any possible disc truncation scenario at lower luminosity or lower Eddington fraction. This paper is organized in the following manner. First, we describe the observations and the details of data reduction in sec .2. In sec. 3 and sec. 4, we describe the temporal and spectral analysis, respectively. Finally, in sec.5, we discuss our findings. 
 
\begin{figure*}
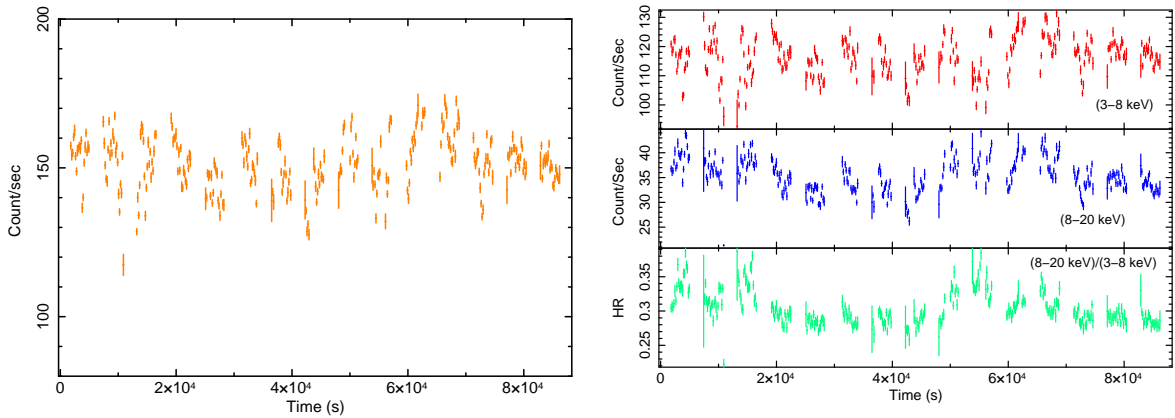

\centering
\includegraphics[scale=0.32, angle=-90]{light_curve.ps}
\includegraphics[scale=0.32, angle=-90]{HR.ps}
\caption{Left: $3-79\kev{}$ \nustar{}/FPMA light curve of Ser~X-1 with a binning of 100 sec. Right: The source count rate in the energy band $3-8$ keV and $8-20$ keV are shown in the the upper and the middle panels, respectively. Bottom panel shows the HR which is the ratio of count rate in the energy band $8-20$ keV and $3-8$ keV.} 
\label{Fig1}
\end{figure*}

\section{observation and data reduction}
After observing twice on 2013 July 12 and 13, \nustar{} observed the source Ser~X-1 again on 2018 February 17 for a total exposure time of $\sim 31$ ks (Obs. ID: $30363001002$). The two observations taken in 2013 have been analyzed by \citet{2013ApJ...779L...2M} and \citet{2017A&A...600A..24M}. \nustar{} data of the source Ser~X-1 were collected with the two co-aligned grazing incidence hard X-ray imaging Focal Plane Module telescopes (FPMA and FPMB) in the $3-79 \kev$ energy band. \\

We reprocessed the data with the standard \nustar{} data analysis software ({\tt NuSTARDAS v1.7.1}) and {\tt CALDB} ($v20181030$). Using the {\tt nupipeline} tool (version v 0.4.6), we filtered the event lists. We used a circular extraction region with a radius of 100 arcsec centered around the source position to produce a source spectrum for both the telescopes, the FPMA and the FPMB. We used another 100 arcsec circular region away from the source position for the purpose of background subtraction. We created lightcurve, spectra and response files for the FPMA and the FPMB using the {\tt nuproducts} tool. We grouped the FPMA and the FPMB spectral data with a minimum of 500 counts per bin (as the source is relatively bright) and fitted the two spectra simultaneously.

\section{Temporal Analysis}
\begin{figure}
\centering
\includegraphics[width=7.0cm, angle=0]{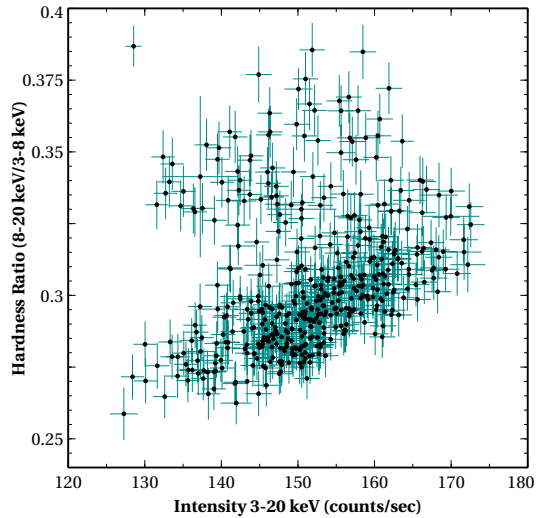}
\caption{The hardness-intensity diagram (HID) of Ser~X-1. Intensity is taken as $3-20$ keV source photon count rate and hardness ratio has been taken as the ratio of photon count rate in the energy band $8-20$ keV and $3-8$ keV. A positive corelation is observed between intensity and the hardness ratio.} 
\label{Fig2}
\end{figure}

Left panel of Figure~\ref{Fig1} shows the $3-79\kev{}$ \nustar{}/FPMA light curve of Ser~X-1 with a binning of 100 sec and spans $\sim31$ ks. Ser~X-1 is a bright X-ray source and in this observation it was detected at an average intensity of $\simeq150$ counts s$^{-1}$. Ser~X-1 is also known for the bursting behaviour but no X-ray bursts were observed during this observation. We also extracted the light curves in the $3-8\kev{}$ and $8-20\kev{}$ energy ranges, with a bin size of 100~s and presented those seperately in the right panel of Figure~\ref{Fig1}. We generated the hardness ratio (HR) between the photon counts in the above mentioned energy bands and displayed it in the right panel of Figure~\ref{Fig1}. The HR value, which is a broad measure of the spectral shape, remained fairly constant with a mean value of $\sim 0.3$. Although  a small count rate variability is observed in the $3-8\kev{}$ and $8-20\kev{}$ energy band, it is not associated with any significant change in the HR. It suggests that the spectral shape of the source remain stable during the whole span of this particular observation. Additionally, we generated the hardness-intensity diagram (HID), in which the HR ($8-20\kev{}$ to $3-8\kev{}$ photon count ratio) is plotted as a function of the source intensity ($3-20\kev{}$), shown in Figure~\ref{Fig2}. In this observation, the HID shows that the HR is positively correlated with intensity for this source. In the case of atoll sources, the positive correlation between the hardness and the intensity is characteristic to the banana branch \citep{1993PASJ...45..801A, 1989A&A...225...79H}. This means that the source remained in the banana branch rather than in the island branch or in any other period of extreme or unusual behaviour during this observation. Previous \nustar{} observation was also found to sample the usual banana branch \citep{2013ApJ...779L...2M}. However, it may be noted that the island state has so far not been observed from Ser~X-1.

\section{spectral analysis}
We fitted both the \nustar{} FPMA and FPMB spectra simultaneously as initial fits revealed a good agreement between these two spectra. An initial inspection of the FPMA and the FPMB spectra also suggests that the source is detected significantly upto $30 \kev{}$. We therefore performed the spectral analysis over the $3-30 \kev{}$ energy band using {\tt XSPEC  v 12.9}. We added a constant between the spectra to account for uncertainties in the flux calibration of the detectors. The constant was set 1 for the FPMA and left it free for the FPMB. A value of $1.02$ was measured for the FPMB. We modelled the interstellar absorption along the line of sight using the {\tt tbabs} model with {\tt vern} cross sections \citep{1996ApJ...465..487V}and {\tt wilm} abundances \citep{2000ApJ...542..914W}. For each fit, we fixed the absorption column density to the \citet{1990ARA&A..28..215D} value of $4.0\times10^{21}$ cm$^{-2}$ as the \nustar{} data only extends down to $3 \kev{}$ and found it difficult to constrain from the spectral fits. All the uncertainties in this paper are quoted at $90\%$ of the confidence level if not stated otherwise in particular.

\begin{figure*}
\includegraphics[scale=0.40, angle=-90]{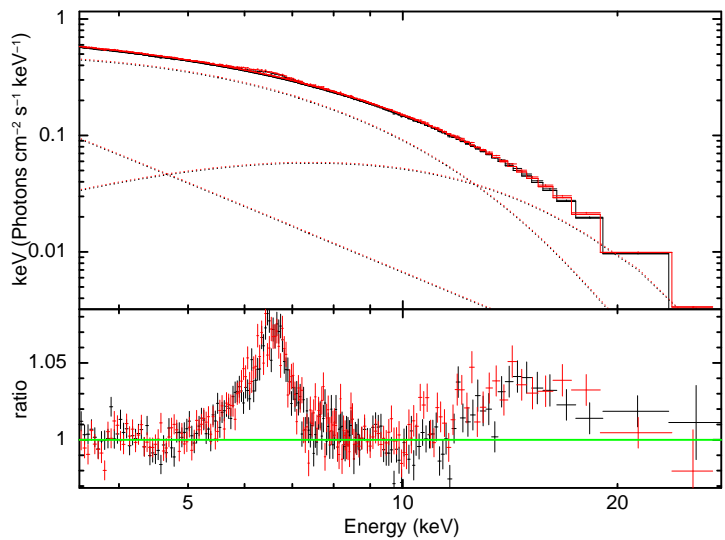}
\caption{\nustar{} (FPMA in black, FPMB in red) unfolded spectra. The data were fit with the model consisting of an absorbed multicolour disk blackbody, a single temperature blackbody and a powerlaw. Model used: {\tt TBabs$\times$(diskbb+bbody+powerlaw)}. It revaled un-modelled broad emission line $\sim 5-8$ keV and a clear hump like feature $\sim 10-20$ keV. The prominent residuals can be indentified as a broad Fe-K emission line and the corresponding Compton back-scattering hump. The spectral data were rebinned for visual clarity} 
  \label{Fig3}
   \end{figure*} 

\subsection{Continuum modeling}
We fitted $3-30 \kev{}$ \nustar{} continuum to a model consisting of a disc blackbody component ({\tt diskbb} in {\tt XSPEC}), a single-temperature blackbody component ({\tt bbody} in {\tt XSPEC}) and a power-law component ({\tt powerlaw} in {\tt XSPEC}). This combination of models can be interpreted in terms of the emission from the accretion disc, the emission likely caused by the boundary layer/NS surface and the hard emission that may arise through the Comptonization of soft photons from high energy (thermal or non-thermal) electrons in the corona. This combination of models describes the shape of the continuum very well although the $\chi^2$ ($\chi^{2}/dof$=$2450/739$) is not acceptable because of the presence of the strong disc reflection features in the spectrum as it is evident in Figure~\ref{Fig3}. We note that, if we eliminate the {\tt powerlaw} component from this continuum model, we get a worse fit, corresponding to a decrease of $\Delta\chi^2=191$ for the addition of two parameters when the {\tt powerlaw} component is included in this fit. So, {\tt powerlaw} component remains significant as it required to fit high-energy residuals of the atoll sources in the soft state (see e.g. \citealt{2015MNRAS.450.2016P, 2001ApJ...548..883I, 2006A&A...459..187P, 2007ApJ...654..494T, 2007A&A...471L..17P, 2006ApJ...651..416F}) and also for the Z-sources (e.g. \citealt{2000ApJ...544L.119D}). This combination of continuum model has been frequently used for the soft state spectra of many atoll type NS LMXBs \citep{2007ApJ...667.1073L, 2010ApJ...720..205C, 2013ApJ...779L...2M}. \\

Emission from the boundary layer can also be modelled via low-temperature, optically thick Comptonization. To test this, we replaced the single-temperature blackbody component by the Comptonization model {\tt nthcomp} \citep{1996MNRAS.283..193Z, 1999MNRAS.309..561Z} in {\tt XSPEC}, setting the seed photon spectral shape to a blackbody. Such replacement gives a similar quality fit ($\chi^{2}/dof$=$2431/738$) in comparison with the simpler continuum model. It may be noted that, the norm of the {\tt powerlaw} component becomes remarkably small ($\sim2.35\times10^{-11}$) which indicates that the {\tt powerlaw} component appears to be much weaker in this case. Moreover, the elimination of the {\tt powerlaw} component from the fit does not lead to any notable change in the value of the $\chi^{2}/dof$. Therefore, we remove this component from this fit as it is not statistically required. As both the models, a simpler continuum {\tt tbabs$\times$(diskbb+bbody+powerlaw)} and a continuum consisting of a Comptonized component and a disc blackbody {\tt tbabs$\times$(diskbb+nthcomp)} represent the continuum fairly well, we proceed with both choices of the continuum models and use those in all the following analyses.

\begin{figure*}
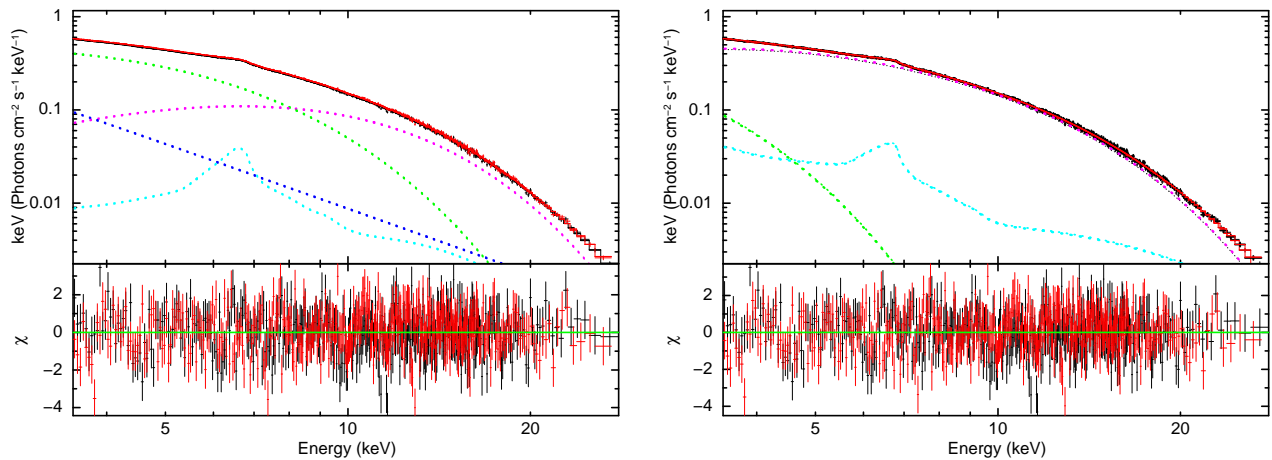

   \includegraphics[scale=0.34, angle=-90]{best_fit_repeat.ps}
   \includegraphics[scale=0.34, angle=-90]{best_fit_nthcomp.ps}
   \caption{The \nustar{} (FPMA in black, FPMB in red) unfolded spectra of Ser~X-1 with the best-fitted models. \textbf{Left:} Model consisting of a disk blackbody (green), a single temperature blackbody (purple), a powerlaw (blue) and a relativistically blurred reflection model (cyan) i.e.,{\tt TBabs$\times$(diskbb+bbody+powerlaw+relconv*reflionx)}. Lower panel shows residuals in units of $\sigma$.  \textbf{Right:} Model where the single temperature blackbody and powerlaw in the previous model is replaced by the Comptonized component {\tt nthcomp}(purple) and the standard {\tt reflionx} model (cyan) is used although modified in order to mimic a Comptonization input spectrum. The overall model now is {\tt TBabs$\times$(diskbb+nthcomp+highecut*rdblur*reflionx).} }  
   \label{Fig4}
   \end{figure*}

\subsection{Reflection Model}
There are two strong emission features around $\sim 5-8\kev{}$ and $10-20\kev{}$ (see Figure~\ref{Fig3}), indicating the presence of a reflection component in the Fe K$\alpha$ region and the corresponding Compton back-scattering hump.
In NS systems, the accretion disc may be illuminated by the thermal emission coming from the boundary layer of the NS or by the Comptonization continuum, resulting in reflected emission. In this observation the continuum of Ser~X-1 is dominated by the blackbody component. We therefore included a modified version of the {\tt reflionx} \footnote{https://www-xray.ast.cam.ac.uk/\~{}mlparker/reflionx\_models \\/reflionx\_bb.mod} model ({\tt reflionx$\_$bb}) that assumes that the disc is illuminated by a blackbody, rather than a power-law (see e.g. \citealt{2010ApJ...720..205C, 2016ApJ...819L..29K, 2016MNRAS.456.4256D}). As the replacement of the hot blackbody component with the Comptonization model provides similar description of the continuum, we have used in this case the standard {\tt reflionx} model although modified in order to mimic a Comptonization input spectrum (see below). The parameters of the {\tt reflionx$\_$bb} model include the disc ionization parameter ($\xi$), the iron abundance ($A_{Fe}$), the temperature of the ionizing black body flux $kT_{refl}$ and a normalization $N_{refl}$. In order to account for relativistic Doppler shifts and the gravitational redshifts, we convolved {\tt reflionx$\_$bb} with {\tt relconv} \citep{2010MNRAS.409.1534D}. Its parameters include the inner and the outer disc emissivity indices ($q_{in}, q_{out}$), break radius ($R_{break}$), the inner and outer disk radii $R_{in}$ and $R_{out}$, the disk inclination ($i$) and the dimensionless spin parameter ($a$). In our fits, we used a constant emissivity index (fixed slope) by fixing $q_{out}=q_{in}$ which essentially obviates the meaning of $R_{break}$.  \\

We imposed a few reasonable conditions when making fits with reflection models. We set the emissivity to $q=3$, in agreement with a Newtonian geometry far from the NS \citep{2010ApJ...720..205C}. We set a redshift of $z=0$ since Ser~X-1 is a Galactic source. We fixed the spin parameter $a=0$ since most NS in LMXBs have $a\le0.3$ \citep{2008ApJS..179..360G, 2011ApJ...731L...7M}. The difference of measurement of the  position of the $R_{ISCO}$ between $a=0$ and $a=0.3$ is less than $1\;R_{g}$ (where $R_{g}=GM/c^2)$. The dimensionless spin parameter $a$ can be approximated as $a\simeq0.47/P_{ms}$ \citep{2000ApJ...531..447B} where $P_{ms}$ is the spin period in ms. The spin period of the source Ser~X-1 is unknown as kHz quasi-periodic oscillations (QPOs) and burst oscillations have not been detected so far. We therefore performed the fit with $a=0$ as well as $a=0.3$ and found that both the fit yielded similar results. The outer disc radius was fixed to $1000\;R_{g}$. In order to get a consistent result, we linked the reflected blackbody temperature $kT_{refl}$ in the {\tt reflionx$\_$bb} model to the temperature of the blackbody component $kT_{bb}$. \\

The addition of the relativistic reflection model improves the spectral fits significantly ($\chi^{2}/dof$=$886/734$). The best-fitting parameters for the continuum and the reflection spectrum are reported in Table~\ref{parameters}. Our fits suggest that the inner disc $R_{in}$ is truncated prior to the ISCO at $(1.9-2.5)\;R_{ISCO}$ ($\simeq11.4-15\,R_{g}\: \text{or}\: 26-34$ km), in agreement with \citet{2017A&A...600A..24M}. The inclination angle is found to be $i\sim18\degr$ in agreement with the fact that neither dips nor eclipses have been observed in the light curve of Ser~X-1. This result is in line with a general agreement that the inclination angle in this source is relatively low ($<50\degr$) as observed by other authors (see e.g. \citealt{2017A&A...600A..24M, 2010ApJ...720..205C, 2018ApJ...858L...5L}), although \citet{2013ApJ...779L...2M} and Optical observations \citep{2013MNRAS.432.1361C} pointed to a very low inclination angle of the system ($\sim10\degr$ or less). The reflection component has an intermediate disc ionization of $\xi\simeq156-248$ erg s$^{-1}$ cm which is consistent with $\text{log}\xi\sim (2-3)$ seen in other NS LMXBs (see e.g. \citealt{2010ApJ...720..205C}). The iron abundance is found to be $A_{Fe}=2.5^{+0.9}_{-0.8}$. The fitted spectrum with relativistically blurred reflection model and the residuals are shown in the left panel of Figure~\ref{Fig4}. \\

\begin{figure*}
\includegraphics[width=6.0cm, angle=0]{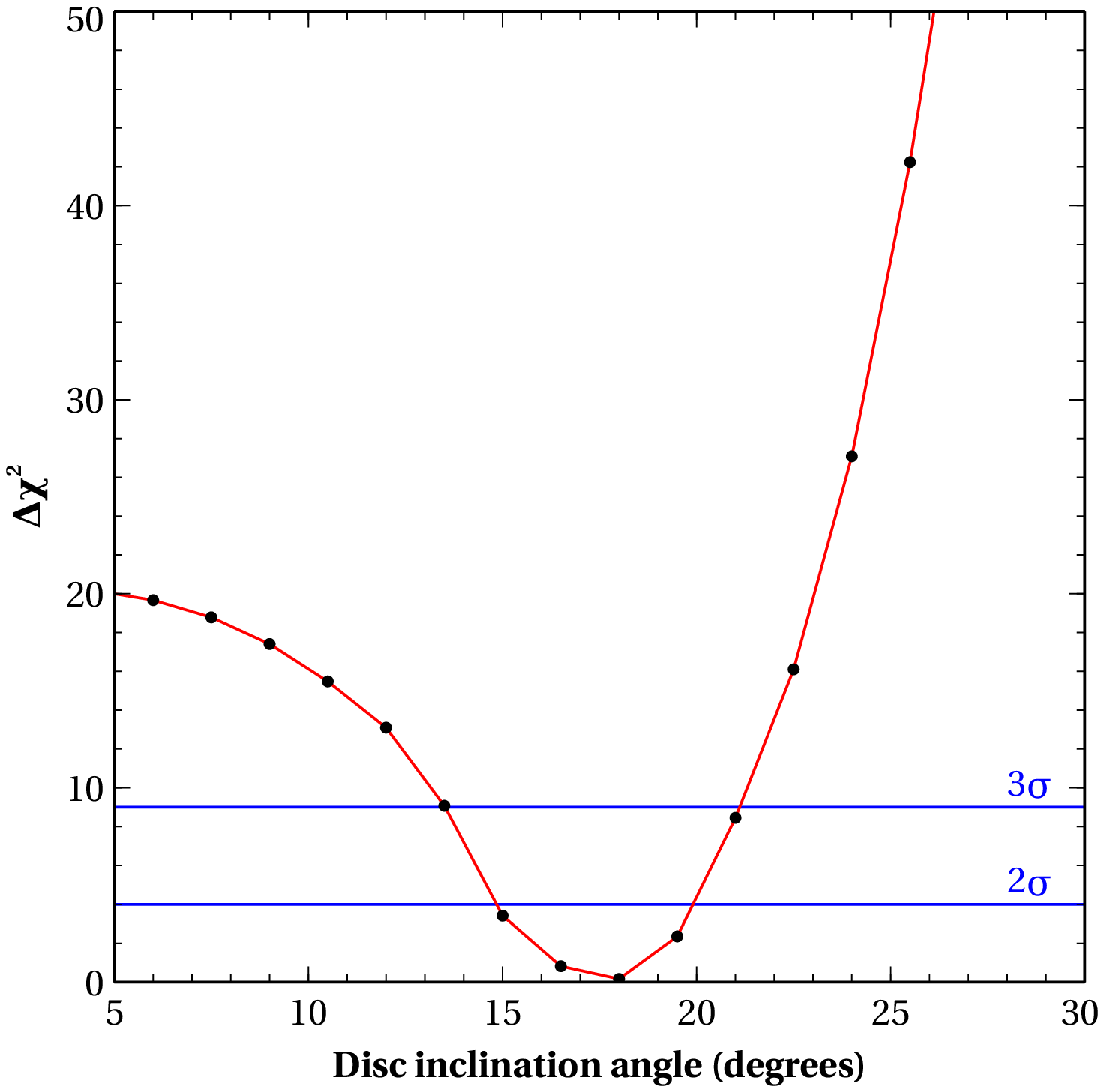}\hspace{1cm}
\includegraphics[width=6.0cm, angle=0]{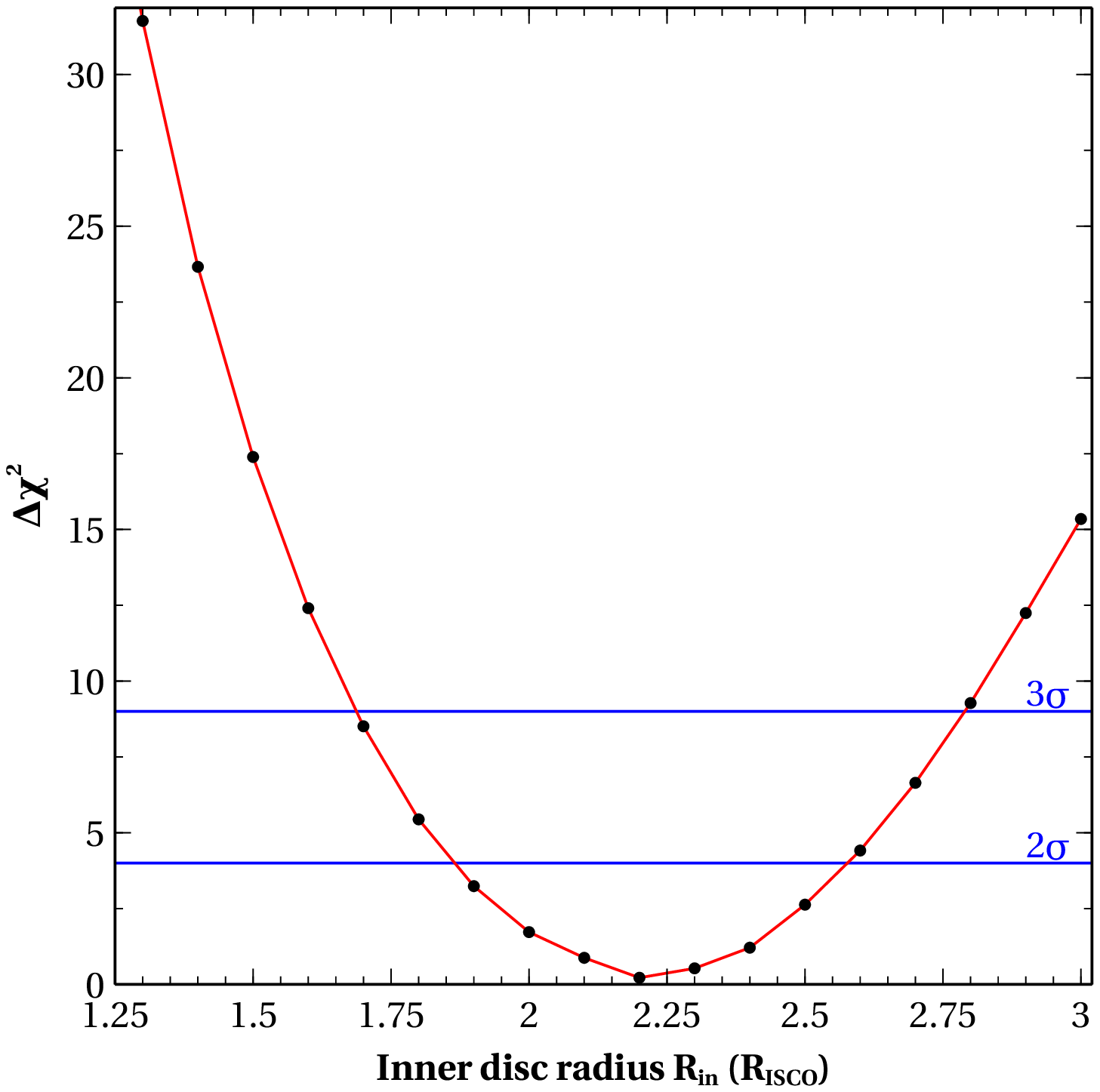}
\caption{Left panel shows the variation of $\Delta\chi^{2}(=\chi^{2}-\chi_{min}^{2})$ as a function of the disk inclination angle obtained from the relativistic reflection model ({\tt reflionx$\_$bb.mod}). We varied the disc inclination angle between 0 degree and 30 degree. Right panel shows the variation of $\Delta\chi^{2}(=\chi^{2}-\chi_{min}^{2})$ as a function of inner disc radius (in units of $R_{ISCO}$) obtained from the relativistic reflection model. We varied the inner disc radius as a free parameter upto $3\,R_{ISCO}$. The value of the inner disc radius is not consistent with the position of the ISCO. Horizontal lines in both the panels indicate $2\sigma$ and $3\sigma$ significance level.} 
\label{Fig5}
\end{figure*}

We computed $\Delta\chi^2$ for the parameters inner disc radius ($R_{in}$) and disc inclination angle ($i$) using {\tt steppar} command in {\tt xspec} to determine how the goodness-of-fit changed as a function of these parameters. The plots of $\Delta\chi^2$ versus $i$ and $R_{in}$ for the best-fit model are shown in the left and right panel of Figure~\ref{Fig5}, respectively. These plots illustrate the sensitivity of the spectra to the inner extent of the disc as well as to the disc inclination angle. This suggests that the inner disc is truncated prior to the ISCO and this result is in agreement with what is found by \citet{2017A&A...600A..24M}. \\

Considering the fact that the reflection parameters are mostly constrained by the iron line, we have attempted to fit the data with a relativistic line profile, {\tt relline} model which excludes the broadband features such as the Compton hump seen around $10-20 \kev{}$. We  measured a line centroid energy of $E=6.89^{+0.03}_{-0.04}$\kev{}. The value for the inner disc radius, $R_{in}=1.76^{+0.15}_{-0.10}\; R_{ISCO}$ is consistent with our above estimation. We have found a small inclination of $i=12\pm3\degree$ which is also comparable to the previous \nustar{} observation. However, the {\tt relline} model, with $\chi^{2}/dof$=$976/735$, is not as good a fit as the broadband reflection model described above ($\chi^{2}/dof$=$886/734$). It suggests that the broadband reflection spectrum does make a significant contribution. It may be noted that the line centroid energy is quite inconsistent with a relatively low ionization parameter ($\xi\simeq156-248$ erg s$^{-1}$ cm) found with the reflection model. This may be caused by the fact that the model {\tt relline} does not contain the iron edge that is also present in the reflection component. The lack of this feature may shift the energy of the line to higher values, and produce different best-fit values of the parameters.  \\

\begin{figure}
\includegraphics[width=6.0cm, angle=0]{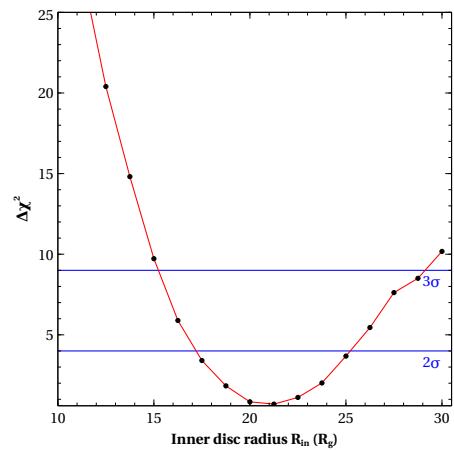}
\caption{Shows the variation of $\Delta\chi^{2}$ as a function of inner disc radius (in units of $R_{g}$) obtained from the  relativistic reflection model ({\tt reflionx.mod}). We varied the inner disc radius as a free parameter between $10\,R_{g}$ and $30\,R_{g}$. Horizontal lines in both the panels indicate $2\sigma$ and $3\sigma$ significance level.} 
\label{Fig6}
\end{figure}

We note that, \citet{2017A&A...600A..24M} re-analysed the same $2013$ \nustar{} dataset (obsID: $30001013002$ and $30001013004$)
analysed by \citet{2013ApJ...779L...2M} using a different choice of the continuum and reflection models. They found that the use of the Comptonization model as a continuum component changed the disc parameters recovered from the reflection component. In the present \nustar{} observation, we have already established that a continuum model consisting of a Comptonized component represents the continuum very well. With this continuum model, we tried to fit the reflection component with the standard {\tt reflionx} \citep{2005MNRAS.358..211R} model that assumes a high energy exponential cutoff power-law irradiating the accretion disc. As cutoff energy of the illuminating power law in the {\tt reflionx} model is set to $300\kev{}$, we have modified {\tt reflionx} in such a manner that it  mimics the nthcomp continuum. In doing so, we  followed exactly the method described by \citet{2017A&A...600A..24M}. In order to consider the high-energy rollover of the Comptonization spectrum, we  multiplied {\tt reflionx} by a high energy cutoff, {\tt highecut} in {\tt XSPEC}, with the low-energy cutoff ($E_{cutoff}$) fixed to $0.1$ \kev. The folding energy $E_{fold}$ was initially set to $\sim3$ times of the electron temperature ($kT_{e}$). We fixed the photon index ($\Gamma$) of the illuminating spectrum to that of the {\tt nthcomp} component. Thus, we have modified the reflection model {\tt reflionx} in order to reproduce the {\tt nthcomp} continuum by introducing the model component {\tt highecut} (for details see \citealt{2017A&A...600A..24M}). To take relativistic blurring into account, we have convolved {\tt reflionx} with {\tt rdblur} component. Since relativistic effects are not extreme for this source, for sake of simplicity we used the {\tt rdblur} component instead of {\tt relconv} in this case. The {\tt rdblur} component depends on the values of the inner and outer disc radii, the inclination angle of the disc and the emissivity index (Betor) which is the index of the power-law dependence of the emissivity of the illuminated disk (scales as $r^{Betor}$). We have performed this fit setting Betor=$-3$, outer disc radius to $1000\,R_{g}$ and $A_{Fe}$ to $1.0$. This has improved the fit significantly to $\chi^{2}/dof$=$892/733$. This fit changes the results for important parameter $R_{in}$ to $22\pm4\,R_{g}$ but inclination remains quite similar ($21^{+1}_{-2}\deg$) to the previous estimation. All the other best-fit parameter values are listed in Table~\ref{parameters01}. The fitted spectrum with this modified relativistically blurred reflection model and the residuals are shown in the right panel of Figure~\ref{Fig4}. We have computed the variation of $\Delta\chi^{2}$ as a function of inner disc radius in between $10\,R_{g}$ and $30\,R_{g}$ and plotted it in Figure~\ref{Fig6}. We have compared the final results of the reflection component when the simpler continuum or a continuum consisting of a Comptonized component are used. \\

Reflection features are better explained by the utilization of the self-consistent reflection model {\tt RELXILL}. Different new flavors of the {\tt RELXILL} model are available today. {\tt RELXILLCP}, a flavor of  {\tt RELXILL}, allows one to use reflection from a Comptonized disc component. This model has a hard-coded seed photon temperature of $0.05$\kev{}. Since we found a higher seed photon temperature of $\sim0.91$\kev{} in our continuum fit with {\tt nthcomp}, we did not attempt to use {\tt RELXILLCP}. Probably this model is not appropriate to describe this particular spectral state. We do not comment on it further.

\begin{table}
   \centering
   \caption{ Best-fitting spectral parameters of the \nustar{} observation of the source Ser~X-1 using model: {\tt TBabs$\times$(diskbb+bbody+powerlaw+relconv*reflionx)}.} 
\begin{tabular}{|p{1.6cm}|p{4.2cm}|p{1.7cm}|}
    \hline
    Component     & Parameter (unit) & Value \\
    \hline
    {\scshape tbabs}    & $N_{H}$($\times 10^{21}\;\text{cm}^{-2}$) &$4.0(f)$     \\
    {\scshape diskbb} & $kT_{disc}$($\kev$) &  $1.70_{-0.07}^{+0.05}$   \\
    & norm [(km/10 kpc)$^{2}\;$cos$i$]   &  $24_{-2}^{+4}$    \\

    {\scshape bbody} & $kT_{bb}(\kev)$ &  $2.35\pm0.04$    \\
    & norm ($\times 10^{-2}$) &  $2.21_{-0.15}^{+0.19}$     \\

    {\scshape powerlaw} & $\Gamma$ &  $3.32\pm0.13$    \\
    & norm    &   $1.81_{-0.37}^{+0.45}$     \\
   
   {\scshape relconv} & $i$ (degrees) &  $18\pm2$    \\
   & $R_{in}$($\times R_{ISCO}$) &  $2.2\pm0.3$ \\
   {\scshape reflionx} & $\xi$(erg cm s$^{-1}$) &  $192_{-36}^{+56}$  \\
   & $kT_{refl}$ (keV) & $2.35\pm0.04 $  \\
   & $A_{Fe}$ ($\times \;\text{solar})$   &  $2.54_{-0.75}^{+0.86} $  \\
   & norm    &   $0.95_{-0.17}^{+0.63}$ \\
   & $F^{*}_{total}$ ($\times 10^{-9}$ ergs/s/cm$^2$) & $5.0\pm 0.01$ \\
   & $F_{diskbb}$ ($\times 10^{-9}$ ergs/s/cm$^2$) &  $2.2 \pm 0.01$  \\
   & $F_{bbody}$ ($\times 10^{-9}$ ergs/s/cm$^2$)&  $2.1 \pm 0.01$ \\
   & $F_{powerlaw}$ ($\times 10^{-9}$ ergs/s/cm$^2$)&  $0.47 \pm 0.01$ \\
   & $F_{reflionx}$ ($\times 10^{-9}$ ergs/s/cm$^2$) &  $0.23 \pm 0.01$  \\
   & $L_{3-30 keV}$ ($\times 10^{38}$ ergs/s) & $0.4 \pm 0.01$  \\	
  
\hline 
    & $\chi^{2}/dof$ & $886/734$   \\
    \hline
  \end{tabular}\label{parameters} \\
{\bf Note:} Here we have used a modified version of the {\tt reflionx} ({\tt reflionx$\_$bb}) that assumes that the disc is illuminated by a blackbody rather than a power-law. The outer radius of the {\tt relconv} spectral component was fixed to $1000\;R_{g}$. The dimensionless spin parameter ($a$) and redshift ($z$) were set to zero. We fixed emissivity index $q=3$, assumed a distance of 7.7 kpc and a mass of $1.5\;M_{\odot}$ for calculating the luminosity. The {\tt diskbb} and {\tt bbody} normalizations imply  radii of $\sim 12$ km and $\sim 5.5$ km respectively, after applying color corrections.\\
$^{*}$All the unabsorbed fluxes are calculated in the energy band $3.0-79.0\kev$. 
\end{table}

  \begin{table}
   \centering
   \caption{ Best-fitting spectral parameters of the \nustar{} observation of the source Ser~X-1 using model: {\tt TBabs$\times$(diskbb+nthcomp+highecut*rdblur*reflionx)}.} 
\begin{tabular}{|p{1.6cm}|p{4.2cm}|p{1.7cm}|}
    \hline
    Component     & Parameter (unit) & Value \\
    \hline
    {\scshape tbabs}    & $N_{H}$($\times 10^{21}\;\text{cm}^{-2}$) &$4.0(f)$     \\
    {\scshape diskbb} & $kT_{disc}$($\kev$) &  $0.67_{-0.19}^{+0.07}$   \\
    & norm [(km/10 kpc)$^{2}\;$cos$i$]   &  $541_{-168}^{+2766}$   \\

    {\scshape nthcomp} & $\Gamma$ &  $2.03\pm0.04$   \\
    & $kT_{e}(\kev)$ &  $2.54\pm0.05$ \\
    & $kT_{bb}(\kev)$ &  $0.89_{-0.11}^{+0.31}$  \\
    & norm  & $0.19\pm0.05$     \\
    {\scshape highecut} & $E_{cut}$($\kev$) & $0.1(f)$    \\
    & $E_{fold}$($\kev$)    &  $10.05_{-2.47}^{+0.95}$     \\

   {\scshape rdblur} & $i$ (degrees) & $21_{-2}^{+1}$    \\
   & $R_{in}$($\times R_{g}$) & $21\pm4$\\
   {\scshape reflionx} & $\xi$(erg cm s$^{-1}$) &  $602_{-280}^{+91}$  \\
   & $\Gamma$  & $2.03\pm0.04 $  \\
   & $A_{Fe}$ ($\times \;\text{solar})$   & $1.0(f)$  \\
   & norm ($\times 10^{-5}$)   &  $1.55_{-0.83}^{+0.92}$ \\
   \hline 
    & $\chi^{2}/dof$ & $891/734$   \\
    \hline
  \end{tabular}\label{parameters01} \\
{\bf Note:} Here we have used the standard {\tt reflionx} model that assumes a high energy exponential cutoff power-law irradiating the accretion disc, modified in such a manner that it  mimics the {\tt nthcomp} continuum (see text).   
The outer radius of the {\tt rdblur} spectral component was fixed to $1000\;R_{g}$. We fixed emissivity index (betor) to $-3$. The $E_{fold}$ parameter is fitted to be 3 times the $kT_{e}$. The {\tt diskbb} normalization implies a radius of $53_{-9}^{+80}$ km after applying color corrections.
\end{table}

\begin{table}
   \centering
   \caption{Luminosity and inner disc radius $(R_{in})$ of previous Ser~X-1 observations.} 
\begin{tabular}{|p{4.5cm}|p{1.2cm}|p{1.5cm}|}
    \hline
    Observation     & L/$L_{Edd}$ & Radius $R_{in}$ $(R_{g})$ \\
    \hline
    {\it XMM}$^1$ (2004, Obs ID 0084020401) & $0.28$ & $25\pm8$\\
    {\it XMM}$^1$ (2004, Obs ID 0084020501) & $0.20$ & $14\pm1$\\ 
    {\it XMM}$^1$ (2004, Obs ID 0084020601) & $0.26$ & $26\pm8$\\
    \suzaku{}$^1$ (2006)   & $0.48$ & $8\pm0.3$ \\[-0.2cm]
    \nustar{}$^2$ (2013)  & $0.62$ &  $7.7\pm0.6$\\[-0.2cm]
    \nustar{}$^3$ (2013)  & $0.62$ &  $13.2\pm3.1$\\
    \chandra{}$^4$ (2014) & $0.38$ & $7.7\pm0.1$\\
    \suzaku{}$^5$ (2014)  & $0.40$ & $8.1\pm0.8$ \\
    \nustar{}$^6$ (2018)  & $0.31$ & $12.9\pm1.7$ \\
\hline
 \end{tabular}\label{parameters1} \\
Information collected from $^1$\citet{2010ApJ...720..205C}, $^2$\citet{2013ApJ...779L...2M}, $^3$\citet{2017A&A...600A..24M}, $^4$\citet{2016ApJ...821..105C}, $^5$\citet{2016ApJ...831...45C}, $^6$present work. The luminosity was calculated based on the $0.5-25 \kev{}$ absorption-corrected flux and a distance of $7.7$ kpc. For more details see Table 4 of \citet{2016ApJ...831...45C}.
 \end{table}

\section{Discussion}
We present here a new \nustar{} observation of the bright atoll type NS LMXB Ser~X-1. The source displayed a particularly soft spectrum with the $3-79\kev{}$ luminosity of $L_{X}\sim0.4\times 10^{38}$ ergs s$^{-1}$ which corresponds to $\sim 23\%$ of the Eddington luminosity assuming a distance of 7.7 kpc. The hardness-intensity diagram suggests that the source was in the so-called banana branch during this observation. The broad-band $3-30\kev$ \nustar{} energy spectrum can be well described either by a three-component continuum model consisting of a disk blackbody ({\tt diskbb}), a single temperature blackbody ({\tt bbody}) and a power-law ({\tt powerlaw}) or by a two-component continuum model consisting of a disk blackbody ({\tt diskbb}) and a  Comptonization spectrum ({\tt nthcomp}). Thermal emission from the accretion disc is prominently detected in the X-ray spectrum. In the X-ray spectrum, we clearly detected a broad iron line $\sim 5-8$ keV and the Compton back-scattering hump peaking at $\sim10-20 \kev{}$ band, indicating the signature of disc reflection phenomenon. The emission from the boundary layer (modelled either by a single temperature hot blackbody or the Comptonization spectrum) provides most of the hard X-ray flux that illuminates the accretion disc and produces the reflection spectrum. Reflection features are best interpreted by a relativistically blurred self-consistent disc reflection model. We have tested different type of reflection models in which the illuminating continuum is different. The shape of the resulting reflection spectrum is determined by the nature of the illuminating continuum that provides the hard X-rays. The major difference between the  reflection models used is that one ({\tt reflionx$\_\text{bb}$}) assumes a blackbody irradiating the disc and another ({\tt reflionx}) using a cutoff power-law input spectrum. Studies on reflection spectra provides valuable insight into the accretion geometry, such as the inner radius and the inclination of the accretion disc and the upper limit on the radius of the NS. \\

From the reflection fit that assumes a blackbody input spectrum, we found that the inner edge of the accretion disc is truncated prior to the ISCO at $R_{in}=(1.9-2.5)\:R_{ISCO}$, given that $R_{ISCO}=6\:GM/c^2$ for an NS spinning at $a=0$. This would correspond to $R_{in}=(11.4-15)\,R_{g}$ or $(26-34)$ km for a $1.5M_{\odot}$ NS. The inner disc radius of the source Ser~X-1 has been measured by different authors \citep{2018ApJ...858L...5L, 2016ApJ...831...45C, 2016ApJ...821..105C, 2007ApJ...664L.103B, 2017A&A...600A..24M, 2010ApJ...720..205C, 2013ApJ...779L...2M} in different flux states using data taken from different satellite missions (see Table~\ref{parameters1}). The inner disc radius spans a range between $\sim8-25\;R_{g}$ and the luminosity between $\sim(3-11)\times 10^{37}$ erg~s$^{-1}$ (corresponding to $L/L_{Edd}\sim0.2-0.6)$. Therefore, our estimated inner disc radius for Ser~X-1 lies within the range obtained from different X-ray missions. The inclination of the system measured from the reflection fit is $\sim18\degr$ which is expected as comparatively low inclination has been reported in a series of past work (\citealt{2017A&A...600A..24M, 2010ApJ...720..205C, 2018ApJ...858L...5L}). From the previous \nustar{} observation, \citet{2013ApJ...779L...2M} derived an inner radius of the disc broadly compatible with the disc extending to the ISCO and an inclination angle $<10\degr$, whereas \citet{2017A&A...600A..24M} re-analyzed the same \nustar{} observation with a different combination of the continuum and reflection model and reported the inner disc radius to be $\sim13\,R_{g}$ with an inclination angle $\sim27\degr$. Therefore, the inferred values of the inner disc radius ($R_{in}$) and the inclination angle ($i$) from our analysis are not as extreme as reported by \citet{2013ApJ...779L...2M} but instead are  more in line with what is reported by \citet{2017A&A...600A..24M}.\\

We found that a continuum model, composed of a disk blackbody ({\tt diskbb}), a single temperature blackbody ({\tt bbody}) and a power-law ({\tt powerlaw}) provided a good description of the spectrum. Additionally, we model the continuum emission with a Comptonized continuum ({\tt nthcomp}) along with a {\tt diskbb}. This combination of the continuum model also represents the spectrum fairly well. Reflection features were prominent for both the continuum models. We applied a different relativistically blurred self-consistent disc reflection model according to the nature of the continuum modeling. We wanted to examine how the choice of continuum and reflection models can affect the important disc parameters ($R_{in}$ and $i$) inferred from the reflection component. For the continuum containing a Comptonization component, we used a modified version of {\tt reflionx} in order to mimic the {\tt nthcomp} continuum by introducing the model component {\tt highecut} (for details see \citealt{2017A&A...600A..24M}). With this choice of continuum and reflection model, we estimated an inner radius of the disc ranging from $17$ to $25\,R_{g}$ and an inclination angle of the system of $\sim21\degr$. We note that the inclination angle is consistent with our previous estimation but we found a higher value of the inner disc radius. The inferred values are also in broad agreement with what is reported by other authors. \\

\begin{figure}
\includegraphics[width=7.0cm, angle=0]{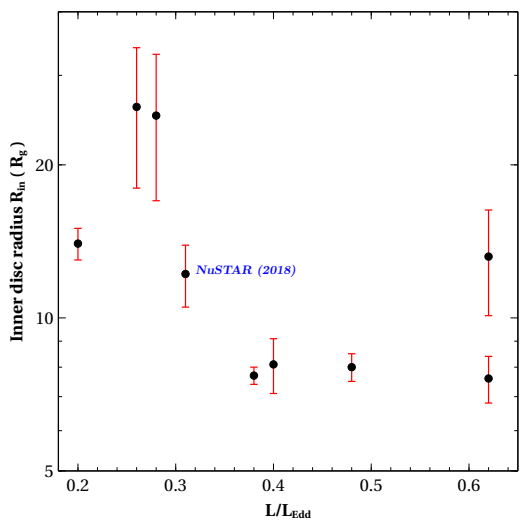}
\caption{Shows the evolution of the inner disc radius $R_{in}$ (in units of $R_{g}$) with the luminosity (based on the data mentioned in Table~\ref{parameters1}). Inner disc radius does not evolve significantly over the range of $L/L_{Edd}\sim0.4-0.6$. A clear trend between $R_{in}$ and $L/L_{Edd}$ is not observed.} 
\label{Fig7}
\end{figure}

In order to see how the inner disc radius evolves with luminosity, we plotted all the inner radius obtained from different observations taken at different flux states and displayed it in Figure~\ref{Fig7}.
It is suggested from Figure~\ref{Fig7} that comparatively larger values of the inner disc radius are obtained in the lower luminosity states when $L/L_{Edd}\sim0.2-0.35$. At the same time, inner disc radius does not appear to change much when $L/L_{Edd}\sim0.4-0.6$ except for the last point which reflects the result of \citet{2017A&A...600A..24M}. So, the accretion disc is likely to be truncated at larger radius at fluxes~$\!\!<0.4 L/L_{Edd}$ (see also \citealt{2016ApJ...831...45C}). This statement is driven by the two higher points with large errors bars and with the exception of the last point. The larger inner radii are mainly obtained from the \xmm{} observations which are less reliable in comparison to other archival data due to short exposure and calibration issues of EPIC-PN timing mode data \citep{2012MNRAS.422.2510W}. From the previous \nustar{} observation, \citet{2013ApJ...779L...2M} found the inner disc radius to be $7.7\pm0.6\;R_{g}$ when the luminosity of the source was $L_{X}\sim1.1\times 10^{38}$ erg~s$^{-1}$ ($\sim 62\%$ of the Eddington luminosity) based on $0.5-40 \kev{}$ flux and assuming a distance of $7.7$ kpc. From this observation, we found an inner disc radius $12.9\pm1.7\;R_{g}$ when the source luminosity was $L_{X}\sim0.54\times 10^{38}$ erg~s$^{-1}$ ($\sim 31\%$ of the Eddington luminosity) based on $0.5-40 \kev{}$ flux and a assuming the same distance of $7.7$ kpc. Moreover, it may be noted that \citet{2017A&A...600A..24M} have re-analyzed the same \nustar{} observation as done by \citet{2013ApJ...779L...2M} with a slightly different continuum and reflection models and found the inner disc radius to be $\sim13\;R_{g}$ which is consistent with our estimation. Therefore, if we compare the inner disc radius obtained by us with that obtained by \citet{2013ApJ...779L...2M}, it seems that the disc appears to move outward during the low luminosity state. But contradiction comes if we compare it with \citet{2017A&A...600A..24M} as our result is consistent with their estimation. So, the commonly assumed fact that the accretion disc in LMXBs moves away from the NS when the X-ray luminosity decays \citep{2017ApJ...847..135L} is not clearly seen based on the \nustar{} observations. This is in agreement with previous results based on the source 4U~1705--44 (see e.g. \citealt{2013A&A...550A...5E, 2015MNRAS.449.2794D}).  \\ 

The ISCO of a gravitating source depends on the mass and radius of the NS. The value of ISCO lie somewhere between $5-6\:R_{g}$ for some reasonable parameters. Our measured inner disc radius of $\sim13\:R_{g}$ is therefore larger than the expected ISCO and the NS surface. The disc could therefore be truncated by either the boundary layer which lies between the disc and the NS surface or by the associated NS magnetic field. From the persistent flux ($F_{p}$) and the distance ($d$) of the source, we estimate the mass accretion rate ($\dot{m}$) per unit area at the NS surface \citep{2008ApJS..179..360G}. Here we used Equation (2) of \citet{2008ApJS..179..360G}
\begin{equation}
\begin{split}
\dot{m}=&\:6.7\times 10^{3}\left(\frac{F_{p}c_\text{bol}}{10^{-9} \text{erg}\: \text{cm}^{-2}\: \text{s}^{-1}}\right) \left(\frac{d}{10 \:\text{kpc}}\right)^{2} \left(\frac{M_\text{NS}}{1.4 M_{\odot}}\right)^{-1}\\
 &\times\left(\frac{1+z}{1.31}\right) \left(\frac{R_\text{NS}}{10\:\text{km}}\right)^{-1} \text{g}\: \text{cm}^{-2}\: \text{s}^{-1},
 \end{split} 
\end{equation}
where $c_{bol}$ is the bolometric correction which is $\sim 1.38$ for the nonpulsing sources \citep{2008ApJS..179..360G}. $M_{NS}$ and $R_{NS}$ are the mass and radius of the NS, respectively. $z$ is the surface redshift and $1+z=1.31$ for a NS with mass 1.4 $M_{\odot}$ and radius 10 km. We determine the mass accretion rate using $F_{p}=5.0\times 10^{-9}$ erg~s$^{-1}$ cm$^{-2}$ to be $\sim5.4\times10^{-9}\;M_{\odot}\;\text{y}^{-1}$ during this observation. Similar mass accretion rates are usually observed when the atoll sources lie in the banana branch.
From the inferred mass accretion rate, we estimate the maximum radial extent ($R_\text{max}$) of the boundary layer region using Equation (2) of \citet{2001ApJ...547..355P}
\begin{equation}
\text{log}(R_\text{max}-R_\text{NS})\simeq5.02+0.245\left|\text{log}\left(\frac{\dot{m}}{10^{-9.85}\:M_{\odot}\:\text{yr}^{-1}}\right)\right|^{2}
\end{equation}
It gives a maximum radial extent of $\sim 6.4\:R_{g}$ for the boundary layer (assuming $M_{NS}=1.5\:M_{\odot}$ and $R_{NS}=10$ km). So, the radial extent of the boundary layer is indeed in agreement with the position of the inner disk. \\

There is also the possibility that the magnetic field is responsible for the disc truncation in this source. If the disc is truncated by the magnetosphere, we can estimate an upper limit on the strength of the magnetic field of the NS using the upper limit of $R_{in}=15\:R_{g}$ measured from the reflection fit. We used Equation (1) of \citet{2009ApJ...694L..21C} to calculate the magnetic dipole moment ($\mu$). Assuming a mass of $1.5\:M_{\odot}$, taking the distance to be $7.7$ kpc and using the unabsorbed flux from $0.01-100 \kev{}$ of $8.5\times10^{-9}$ erg~cm$^{-2}$ s$^{-1}$ as the bolometric flux ($F_{bol}$), we determine $\mu\sim9.3\times10^{26}$ G~cm$^3$ (assuming $k_{A}=1$ which is a factor depending on the geometry, spherical or disk-like, of the accretion flow, $f_{ang}=1$ which is known as the anisotrophy correction factor and accretion efficiency in the Schwarzschild metric $\eta=0.1$). This corresponds to a magnetic field strength of $B\leq1.9\times10^{9}$ G at the magnetic poles for a NS of radius 10 km. This order of magnetic field has the potential to truncate an accretion disc far from the stellar surface \citep{2015MNRAS.452.3994M}. NSs can produce coherent pulsations from the polar caps if the magnetic field truncates the disc. Although, coherent pulsations have never been observed in this source so far, the source could still be magnetically accreting. Pulsations would be undetectable if either the hot spot is nearly aligned with the spin axis or the modulated emission is scattered by the circumstellar gas. For a comprehensive discussion over the issue we refer to \citet{1985Natur.317..681L}.  \\

Another possibility of disc truncation may be due to the state transition associated with a receding disc which is related with a low luminosity and hard power-law dominated X-ray spectra \citep{1997ApJ...489..865E}. The Ser~X-1 spectra presented here are taken in a soft, high luminosity state. Moreover, no state transition has been observed from the HR or HID. Significant disc truncation only occurs at a high enough magnetic field and low mass accretion rate, although there are examples where disc truncation occurs at higher luminosities too \citep{2016ApJ...819L..29K}. Therefore, a significant disc truncation scenario can only be tested if the source Ser~X-1 is observed in a low luminosity and hard spectral state. So, the evolution of the inner disc radius remains unclear as the source has only been observed in the high luminosity and soft spectral state.  

\section{Acknowledgements}
We thank the anonymous referee for their thoughtful comments which have improved the quality of the paper considerably.
This research has made use of data and/or software provided by the High Energy Astrophysics Science Archive Research Centre (HEASARC). This research also has made use of the \nustar{} data analysis software ({\tt NuSTARDAS}) jointly developed by the ASI science center (ASDC, Italy) and the California Institute of Technology (Caltech, USA). ASM  and BR like to thank Inter-University Centre for Astronomy and Astrophysics (IUCAA) for their hospitality and facilities extended to them under their Visiting Associate Programme.

\def\apj{ApJ}
\def\apjl{ApJl}
\def\pasp{PASP} \def\mnras{MNRAS} \def\aap{A\&A} \def\physerp{PhR} \def\apjs{ApJS} \def\pasa{PASA}
\def\pasj{PASJ} \def\nat{Nature} \def\memsai{MmSAI} \def\araa{ARAA} \def\iaucirc{IAUC} \def\aj{AJ} \def\aaps{A\&AS}
\bibliographystyle{mn2e}
\bibliography{aditya}

\end{document}